\documentclass[article,11pt]{elsarticle}

\usepackage{tabularx}
\newcolumntype{x}[1]{>{\centering\arraybackslash\hspace{0pt}}p{#1}}

\usepackage{multirow}
\usepackage{multicol}

\usepackage{enumitem}

\usepackage{graphicx}
\usepackage{amsmath}
\usepackage{booktabs}
\usepackage{gensymb}
\usepackage[section]{placeins}

\usepackage{tikz}
\usetikzlibrary{calc,shapes,arrows,fit}

\usepackage{pgfplots}
\usepackage{pgfplots}
\usepgfplotslibrary{polar}
\usepackage[title]{appendix}
\usepackage{verbatim}

\usepackage{multirow}
\usepackage{soul}
\usepackage{ulem}
\usepackage{pdflscape}

\usepackage{placeins}
\usepackage{amssymb}

\usepackage[margin=0.9in]{geometry}
\usepackage{float}
\usepackage{hyperref}
\hypersetup{
    colorlinks=true,
    linkcolor=blue,
    citecolor=red,
    filecolor=magenta,      
    urlcolor=cyan,
}

\usepackage{nomencl}
\makenomenclature

\journal{TBD}

\usepackage{longtable}
\usepackage{xcolor}
\usepackage{array}
\usepackage{ragged2e}
\newcolumntype{P}[1]{>{\RaggedRight\hspace{0pt}}p{#1}}

\usepackage[font={color=blue,footnotesize},figurename=Figure,labelfont={bf}]{caption}

\usepackage{lineno}
\biboptions{sort&compress}
\usepackage{parskip}
    \usepackage[skip=0.1ex, belowskip=-2ex, margin=-.5cm,
                singlelinecheck=off,labelfont = normalfont]{subcaption}

\begin{document}

\begin{frontmatter}

\title{Draft Version 5 \\ \vspace{20pt} \LARGE{\textbf{Modeling conflicting incentives in engineering senior capstone projects: A multi-player game theory approach}}}

\author{\large Richard Q. Blackwell$^{1}$, Eman Hammad$^{2}$, Congrui Jin$^{2}$, Jisoo Park$^{2}$, Albert E. Patterson$^{2*}$}

\address{\scriptsize
$^{1}$University of Illinois at Urbana-Champaign, Urbana, Illinois, USA  \\
$^{2}$Texas A\&M University, College Station, Texas, USA  \\
\vspace{6pt} $^*$Correspondence: Albert E. Patterson at \url{aepatterson5@tamu.edu}}

\begin{abstract}
University engineering capstone projects involve sustained interaction among students, faculty, and industry sponsors whose objectives are only partially aligned. While capstones are widely used in engineering education, existing analyses typically treat stakeholder behavior informally or descriptively, leaving incentive conflicts, information asymmetries, and strategic dependencies underexplored. This paper develops a formal game-theoretic framework that models capstone projects as a sequential Bayesian game involving three players: (1) The university, (2) the industry sponsor, and (3) the student team. The framework is intended as an analytical and explanatory tool for understanding how institutional policy choices, such as grading structures, intellectual property rules, and sponsor engagement expectations, shape stakeholder behavior and project outcomes, rather than as a calibrated or predictive model. The university acts as a constrained Stackelberg leader by committing to course policies and assessment structures, anticipating strategic responses by sponsors and students under incomplete information. Reduced form outcome functions capture technical quality, documentation quality, timeliness, alignment with sponsor needs, and publishability, while payoff functions reflect stakeholder-specific objectives and costs. Under standard assumptions, the model admits stable equilibrium regimes that correspond to empirically recognizable capstone dynamics observed in practice. These include cooperative engagement, sponsor-dominated exploitation, and student grade gaming. Rather than claiming precise prediction, the framework is intended as an analytical tool for reasoning about incentive design, policy tradeoffs, and structural failure modes in project-based learning environments. The results provide a rigorous foundation for comparative analysis of capstone structures and for future extensions incorporating richer dynamics, repeated interaction, and empirical calibration.
\end{abstract}


\end{frontmatter}





\section{Introduction}
\label{Sec1}
University engineering capstone projects occupy a unique position at the intersection of education, applied engineering practice, and industry collaboration~\cite{ward2013, ritenour2020, farr2001}. They are designed to function simultaneously as culminating academic experiences, structured learning environments, and authentic opportunities for students to engage with real stakeholders. However, these projects exist within a complex ecosystem of incentives that shape the behavior of the three primary actors involved~\cite{rawal2021}: The student team, the sponsoring company and the university that organizes and supervises the course~\cite{watkins2011, rawal2021, howe2019}. Although often assumed to be cooperative and mutually beneficial, these relationships frequently involve misaligned interests, conflicting priorities, and asymmetric expectations. Understanding these tensions is essential for designing sustainable, equitable, and educationally effective capstone programs. In practice, each player enters the capstone ecosystem with its own goals~\cite{howe2019, nassersharif}. Students seek to maximize learning, earn strong grades, build interesting work portfolios, and cultivate professional connections, while managing the workload imposed by competing courses and limited time. Universities, meanwhile, are responsible for ensuring that capstone work satisfies accreditation requirements, provides measurable learning outcomes, upholds academic standards, and maintains ethical, legal, and safety compliance~\cite{zheng2021, nassersharif}. A positive reputation with both students and industry partners depends on delivering a high-quality educational experience while maintaining long-term, credible relationships with sponsors. Sponsors, on the other hand, may hope to gain usable prototypes, exploratory analysis, or innovative concepts at minimal cost. Some sponsors view capstones primarily as educational outreach and recruiting pipelines, whereas others treat them implicitly as opportunities for low-cost engineering labor.

These divergent objectives create predictable conflict points. Students tend to prefer novel or resume-enhancing project directions that may not align with a sponsor’s practical needs. Sponsors may have business timelines or expectations that do not fit academic calendars. Universities must enforce the documentation, reflection, and assessment requirements that students and sponsors may perceive as overhead. Intellectual property restrictions can limit students’ ability to showcase their work, while unclear mentorship expectations can leave sponsors over- or under-involved. In some cases, each player can “win” only by disadvantaging another~\cite{prince2006, biggs1996, gibbs1999}. For example, when students minimize effort to satisfy only the grading rubric, or when a sponsor extracts value without providing adequate mentorship, or when universities prioritize assessment structures that burden students and diminish sponsor value. This inherent multi-agent tension makes capstone projects well suited for game-theoretic interpretation. The capstone ecosystem can be modeled as a three-player strategic interaction in which the university sets the rules and constraints, the sponsor chooses a stance ranging from supportive to exploitative, and students allocate effort toward academic outcomes or sponsor-driven deliverables. Each player operates with incomplete information about the true priorities of the others, and interactions repeat over semesters, creating reputational and relational feedback loops. Such a model enables systematic exploration of equilibrium behaviors, incentive misalignments, and the conditions under which cooperative or exploitative outcomes emerge.

This paper proposes a conceptual game-theoretic framework for analyzing capstone stakeholder dynamics, highlighting how educational goals, industry expectations, and student motivations intersect. The contribution of this work lies in making the structure of incentives and decision sequencing in capstone projects explicit, rather than in proposing specific functional forms or calibrated parameter values. The framework is designed to explain how stable patterns of behavior can emerge from rational stakeholder responses to institutional policies and engagement expectations. By modeling the university as a first mover that commits to observable course policies, and by characterizing how sponsors and students respond under incomplete information, the analysis provides a structured way to reason about incentive dominance, policy tradeoffs, and recurring capstone dynamics. The goal is not to forecast outcomes for a particular program, but to offer a diagnostic lens for interpreting why certain capstone regimes repeatedly arise across institutions. By examining the incentives, strategies, and potential equilibria between the three players, this framework offers a structured lens to understand both the successes and the recurring dysfunctions of university–industry capstone collaborations. Section~\ref{Sec2} explores some related work that has already been done, helping to motivate the framework development in Section~\ref{Sec3}. Three illustrative case studies are presented in Section~\ref{Sec4}, the lessons from which are discussed in detail in Section~\ref{Sec5}. Finally, Section~\ref{Sec6} offers some concluding remarks on the work in general.

\section{Related Work}
\label{Sec2}
Industry-sponsored capstone projects have been widely studied in engineering education as a mechanism for integrating design, professional practice, and experiential learning~\cite{prince2006}. Early surveys and reviews document the prevalence of capstone courses, their structural diversity, and persistent challenges associated with the balance of educational objectives and sponsor expectations \cite{dutson1997, todd1995, howe2006, howe2010, badir2023}. Despite this extensive descriptive literature~\cite{zheng2021}, most previous work treats stakeholder behavior informally, focusing on course organization, assessment practices, or reported outcomes rather than on the strategic interactions that lead to recurring success and failure modes~\cite{morsy2024}. A parallel body of work in higher education research emphasizes the role of assessment structures in shaping student behavior. Studies on constructive alignment and assessment-driven learning show that students respond rationally to grading incentives, often prioritizing rubric-visible outputs over deeper conceptual understanding when assessment signals are strong \cite{biggs1996, gibbs1999}. This literature provides an important context for grade-gaming behavior observed in capstone environments, but typically lacks a formal framework to analyze how such behavior emerges from interactions among multiple stakeholders~\cite{suleiman2025}. Formal models of strategic interaction, including Stackelberg games and principal-agent frameworks, offer tools for analyzing systems in which an actor commits to policies anticipating rational responses by others \cite{fudenberg1991, bolton2005}. Although these approaches are well developed in economics and organizational theory, their application to engineering education and capstone design has been limited. Existing educational models rarely incorporate incomplete information, strategic response, or equilibrium reasoning among institutional, industrial, and student actors. More broadly, systems-oriented perspectives on education highlight the importance of incentive alignment and institutional structure in shaping behavior \cite{simon1996, scott2014}. These perspectives motivate the present work, which treats capstone projects as strategic systems rather than isolated pedagogical interventions. By integrating insights from engineering education, assessment theory, and game theory, the framework developed here complements existing descriptive studies while providing a formal lens for reasoning about incentive dominance, policy tradeoffs, and persistent capstone pathologies.

\section{Game Theory Framework for Stakeholder Incentives}
\label{Sec3}
This work develops a formal game-theoretic model of the interactions among the three primary stakeholders in a university capstone project: (1) The university, (2) the industry sponsor, and (3) the student team completing the project. The framework is designed to capture the incentive structures, conflicting objectives, incomplete information, and sequential dependencies that naturally arise during capstone projects. By formulating the system as a dynamic Bayesian game, the model enables an examination of how policy choices, sponsor behavior, and student effort allocation jointly shape educational and technical outcomes. 

\subsection{Players and order of moves}
Capstone projects have a hierarchical decision structure. For the purposes of this model, it will be assumed that the university first specifies course policies, grading rubrics, and intellectual property rules. In this baseline model, the university represents the program-level institutional actor that sets course policies and assessment structures. The sponsor then selects a project stance and a level of oversight consistent with these constraints. Finally, the student team allocates its limited effort in response to both university policies and sponsor expectations. This assumed sequence of events is best modeled as a sequential extensive-form game. In this game, there are three players: (1) The university $U$, (2) the sponsor $C$, and (3) the student team $S$. In this formulation, the move order is
\begin{equation}
U \rightarrow C \rightarrow S \rightarrow \text{Outcomes} \rightarrow \text{Payoffs}.
\end{equation}
This formulation is a sequential game with incomplete information and asymmetric observation. The mechanism designer is the entity responsible for specifying the institutional environment in which strategic interactions occur. In this study, the university (through its course policies, assessment structure, and supervisory practices) designs the mechanism that determines how sponsor engagement, student effort, documentation quality, and technical performance translate into utilities. The university acts as a constrained mechanism designer, selecting policy instruments rather than full contingent contracts or monetary transfers. Its role is analogous to that of a principal in principal–agent theory. It cannot directly command stakeholder behavior, but it can structure incentives to encourage desirable equilibrium outcomes. All university policies will be assumed to be publicly observable before the sponsor acts. The sponsor actions are observed by students, but are not fully observed by the university until after the project is completed, reflecting limited mid-semester oversight. The students’ choices are private to the team during the project, but their results enter the observable outcome variables at the end of the term.

\subsection{Type spaces (nature of incomplete information)}
Each player possesses private characteristics (types) that influence strategic behavior, but are not directly observable by the other players. Universities vary in pedagogical orientation and academic priorities; sponsors differ in their willingness to support student learning and dissemination; and students vary in their intrinsic cost of exerting effort. Explicitly modeling these types allows beliefs and signaling to influence equilibrium play. The university type is
\begin{equation}
\theta_U \in \{\theta_U^{L},\,\theta_U^{H}\},
\end{equation}
representing different institutional priorities (e.g., low versus high emphasis on academic rigor, documentation, and publishability). The university type $\theta_U$ captures heterogeneity in institutional priorities rather than strategic capability. In particular, $\theta_U$ parameterizes how the university values educational outcomes relative to administrative costs. This type is not directly observable by other players, but influences how university policies translate into utility. Although the university chooses observable policies $(r,i,m)$ strategically, its underlying orientation $\theta_U$ affects how these policies translate into payoffs. The sponsor has a continuous type
\begin{equation}
\theta_C \in [0,1],
\end{equation}
capturing its underlying orientation towards academic goals (higher $\theta_C$ indicates a more academically supportive organization). In addition to its type, the sponsor chooses a discrete posture $s\in\{\text{supportive},\text{exploitative}\}$, which reflects how it actually engages with the project during the semester. The posture of the sponsor $s$ represents short-term strategic behavior, while $\theta_C$ captures long-term organizational orientation; the two do not need to coincide. The posture is a strategic choice and not a direct revelation of type: High-type sponsors are more likely (but not required) to adopt supportive behavior, and low-type sponsors can still behave cooperatively. The continuous type $\theta_C$ enters the payoff and outcome functions directly, while $s$ captures coarse behavioral modes that depend on both the type and the incentives. Students have heterogeneous effort costs, with type
\begin{equation}
\theta_S \in \{\ell, h\},
\qquad
0 < k(\ell) < k(h),
\end{equation}
where $k(\theta_S)$ determines the marginal cost of effort ($\ell =$ low-cost, $h =$ high-cost). The cost of effort, regardless of how high or low, is greater than zero. The common priors over types are denoted by $p_U(\theta_U)$, $f_C(\theta_C)$, and $p_S(\theta_S)$. These represent the shared initial uncertainty about each stakeholder’s type before any actions are observed. They summarize what players believe about the others before taking actions. As the game progresses, players update these beliefs using Bayes’ rule whenever actions reveal information about the underlying types.

\subsection{Action spaces}
Each stakeholder selects actions consistent with its role in the capstone ecosystem which benefit themselves. The university determines formal policies and expectations for the academic project requirements, the sponsor sets technical expectations and provides oversight, and the students choose the intensity and strategic focus of the effort. The actions of each player are as follows. 

\begin{itemize}
\item \textbf{University action}: The university selects
\begin{equation}
a_U = (r, i, m),
\end{equation}
where $r \in [0,1]$ denotes the strictness of the rubric or academic expectations for the project, $( i \in \mathcal{I}$ specifies the IP/publication policy, and \( m \in \{0,1,2\} \) defines the minimum intensity of mentoring that the university expects the sponsor to provide to the students. The mentoring requirement $m$ is modeled discretely to reflect that universities typically specify engagement in qualitative tiers (for example, hands-off, periodic check-ins, or high involvement) rather than as a continuous quantity. 
\item \textbf{Sponsor action}: The sponsor selects
\begin{equation}
a_C = (s, o, d),
\end{equation}
where $s \in \{\text{supportive},\ \text{exploitative}\}$ is the posture, $o \in \{0,1,2\}$ is the intensity of the mentoring, and $d \in [0,\bar{d}]$ for some finite $\bar{d}$ is the strictness of the project/scope. For the intensity of the mentoring, 0 corresponds to a hands-off sponsor that does not engage with students meaningfully, 1 indicates that the sponsor regularly checks in and has design reviews with the students, and 2 means that the sponsor is actively mentoring the students and providing regular technical support. Mentoring intensity is modeled as a discrete variable because sponsor engagement typically occurs in qualitatively distinct modes (hands-off, periodic engagement, or high involvement), while scope strictness varies more continuously across projects and is therefore represented as a nonnegative real value. The mentoring requirement $m$ is not directly enforceable during the semester and therefore does not constrain the feasible choice of the sponsor of $o$, but enters the university’s return through monitoring costs.  
\item \textbf{Student action}: The students choose
\begin{equation}
a_S = (e, x),
\end{equation}
with effort \( e \in [0,1] \) representing the normalized fraction of total available working time that the team allocates to the project, and strategic orientation \( x \in \{L,D,M\} \), corresponding to a learning-focused approach (L), a deliverable-focused approach (D), or a minimalist effort allocation strategy (M). The orientation variable is modeled discretely to capture qualitatively distinct behavioral modes frequently observed in capstone teams.
\end{itemize}

\subsection{Outcome functions}
The joint actions of all players determine the observable outcomes of the project. For notational simplicity, each outcome is modeled as an expected value conditioned on the relevant actions and, where appropriate, stakeholder types. These outcomes include technical quality, documentation quality, timeliness, alignment with sponsor expectations, and publishability. In this model, publishability refers to the probability that project outcomes lead to academic publications that primarily benefit the university. These outcome functions formalize how incentives translate into actual performance in the capstone project. Here, “project performance” refers to the quality results realized from the final project, not to the utility of the stakeholders or the intermediate behavior during the project. All outcome functions are modeled as linear (affine) mappings of the relevant actions and types. This choice provides transparent sensitivity analyzes, monotonicity of effects, and interpretability of coefficients, and is standard in reduced form representations of complex educational or organizational interactions. The model is not intended to capture detailed production-function nonlinearities but instead to encode the directional effects of actions on observable project outcomes. In this formulation, let the outcome functions be

\begin{itemize}
\item $Q = \text{technical quality, where}$
\begin{equation}
\label{8}
\mathbb{E}[Q \mid e,o,x,s,d]
  = q_0 
    + q_1 e
    + q_2 o
    - q_3 1[s = \text{exploitative}]
    - q_4 d
    + q_5 \, 1[x = L] \cdot o.
\end{equation}
This function states that technical quality increases with student effort $e$ and sponsor mentoring intensity $o$, but decreases when the sponsor adopts an exploitative posture $(s=\text{exploitative})$ or when the scope of the project is overly strict $(d)$. The interaction term $q_5\,1[x=L]\,o$ captures the idea that a learning-focused orientation $(x=L)$ improves technical quality primarily when meaningful mentoring is present. In this specification, learning-oriented teams benefit from guidance that helps translate conceptual understanding into effective engineering output, whereas unsupported learning-focused work may not improve short-term technical deliverables. The coefficients $q_0,\dots,q_5$ represent the baseline quality levels and the marginal effects of each factor.

\item $D_{\text{doc}} = \text{documentation quality, where}$
\begin{equation}
\label{9}
\mathbb{E}[D_{\text{doc}} \mid e,x,r,o]
  = d_0 + d_1 e 
  + d_2 \mathbf{1}[x = L]
  + d_3 r 
  + d_4 o .
\end{equation}
Documentation quality increases with student effort $e$, with a learning-focused orientation $(x = L)$, and with stricter university rubrics $r$, reflecting the institution’s emphasis on reporting, reflection, and process transparency. The positive coefficient on the intensity of mentoring $o$ indicates that the participation of sponsors can improve documentation by providing students with clearer technical expectations, structured feedback, and guidance on how to communicate engineering decisions effectively. Parameters $d_0,\ldots,d_4$ determine the baseline level of documentation quality and the marginal influence of each factor.

\item $T = \text{timeliness, where}$
\begin{equation}
\label{10}
\mathbb{E}[T \mid e,o,d]
  = t_0 + t_1 e + t_2 o - t_3 d.
\end{equation}
The timeliness improves with greater student effort $e$ and greater intensity of sponsor mentoring $o$, both of which help teams maintain steady progress and efficiently resolve technical problems. In contrast, timeliness decreases with stricter scope requirements $d$, as more demanding project requirements take longer to complete. The coefficients $t_0,\dots,t_3$ define the baseline level of timeliness and the marginal effect of each factor in this reduced form representation.

\item $A = \text{sponsor alignment, where}$
\begin{equation}
\label{11}
\mathbb{E}[A \mid x,s,d]
  = a_0 
  + a_1 \mathbf{1}[x = D] 
  + a_2 \mathbf{1}[s = \text{supportive}]
  - a_3 d.
\end{equation}
Sponsor alignment is highest when students adopt a deliverable-focused orientation $(x = D)$, which prioritizes producing output that meets sponsor expectations. Alignment is also higher when the sponsor supports $(s=\text{supportive})$, while exploitative sponsors do not contribute positively to alignment. Alignment decreases with stricter project scopes $d$, which raise expectations and increase the risk of falling short. The parameters $a_0,\ldots,a_3$ encode these relationships.

\item $P = \text{publishability, where}$
\begin{equation}
\label{12}
\mathbb{E}[P \mid i,\theta_C,Q,D_{\mathrm{doc}}]
  = p_0
    + p_1 1[i = \text{permissive}]
    + p_2 \theta_C
    + p_3 Q
    + p_4 D_{\mathrm{doc}},
\end{equation}
with parameters chosen such that $\mathbb{E}[P] \in [0,1]$. Publishability is the expected likelihood that the project produces results that can be published as academic documents such as conference papers, patents, technical reports, and similar. This increases when the university adopts a permissive IP policy $(i=\text{permissive})$, when the sponsor’s type $\theta_C$ is more academically supportive, and when the project shows higher technical quality $Q$ and higher documentation quality $D_{\text{doc}}$. These latter effects capture the idea that projects with sound engineering results and clear communication of methods are more likely to produce publishable results. The parameters $p_0,\ldots,p_4$ are chosen so that the expected publishability is within the probability range $[0,1]$. The coefficients are restricted such that $p_0 + p_1 + p_2 + p_3 + p_4 \leq 1$ and all inputs are normalized to $[0,1]$.

\end{itemize}

\noindent For Equations~\ref{8}, \ref{9}, \ref{11}, and \ref{12}, $\mathbf{1}[\cdot]$ denotes an indicator function that takes the value $1$ when the condition inside the brackets is true and $0$ otherwise. Indicator terms allow the model to capture qualitative shifts in outcomes that depend on categorical choices, such as student orientation or sponsor posture.

\subsection{Stakeholder payoff functions}
Each stakeholder evaluates the project through a payoff function derived from the observable outcomes. These payoffs determine strategic incentives and are used in the equilibrium analysis.

\begin{itemize}
\item \textbf{University payoff}: The university values technical learning, documentation quality, and publishability, while incurring administrative costs associated with enforcing strict rubrics and mentoring requirements
\begin{equation}
U_U = \alpha_Q(\theta_U) Q + \alpha_D(\theta_U) D_{\text{doc}}
      + \alpha_P(\theta_U) P - c_r r - c_m m.
\end{equation}
The weight functions $\alpha_Q(\theta_U)$, $\alpha_D(\theta_U)$, and $\alpha_P(\theta_U)$ are nonnegative and capture differences in institutional priorities between university types. For example, a research-oriented university places greater weight on publishability, while a teaching-oriented institution prioritizes learning and documentation outcomes. These components represent the core educational objectives and the university’s broader mission to generate ``publishable'' student work. The costs $c_r r$ and $c_m m$ reflect administrative burdens associated with enforcing strict grading rubrics and mandating minimum levels of sponsor mentoring. Higher rubric strictness $r$ requires more faculty oversight and evaluation effort, while higher mandated mentoring levels $m$ require monitoring and compliance enforcement. Thus, $U_U$ balances the educational benefits of strong student outcomes with the administrative and operational costs of managing the capstone program.

\item \textbf{Sponsor payoff}: Sponsors value timely progress and alignment with their expectations, while experiencing disutility from providing intensive mentoring
\begin{equation}
U_C = \beta_A A + \beta_T T - c_o o - c_d d^2.
\end{equation}
where $\beta_A, \beta_T$ are nonnegative weights and $c_o, c_d$ are cost coefficients. Sponsor utility increases with higher alignment $A$ and greater timeliness $T$, reflecting the sponsor’s preference for reliable project progress and outcomes that match their expectations. Sponsors incur a linear cost $c_o o$ from providing the mentoring effort $o$, which requires time and personnel resources. The cost of project scope strictness $d$ enters quadratically as $c_d d^2$, capturing the idea that moderately challenging scopes are valuable, but extremely strict or demanding project requirements impose increasing internal burdens on the sponsor, such as the need for additional reviews, oversight, and technical support. These assumptions ensure an interior optimal scope choice and avoid corner solutions in which sponsors would always select minimal scope.

\item \textbf{Student payoff}: Students receive utility from grades, represented here as a weighted combination of observable project outcomes, which depend on technical, documentation, and timeliness outcomes, minus the disutility of effort
\begin{equation}
U_S = \gamma_Q Q + \gamma_D D_{\text{doc}} + \gamma_T T - k(\theta_S)e,
\end{equation}
with $k(\theta_S)$ capturing differences in students' marginal effort costs. Student utility is modeled as a reduced form representation of grade-related benefits and the disutility of effort. In this reduced form specification, grades are represented as a linear weighting of technical quality $Q$, documentation quality $D_{\text{doc}}$, and timeliness $T$ through the coefficients $(\gamma_Q,\gamma_D,\gamma_T)$. The terms $\gamma_Q Q$, $\gamma_D D_{\text{doc}}$, and $\gamma_T T$ capture the idea that student grades depend on technical quality, documentation quality, and timely project completion, with
weights reflecting the relative importance of each dimension in the course assessment. The effort cost $k(\theta_S)e$ represents the disutility of exerting effort, where $k(\theta_S)$ captures heterogeneity in student's marginal effort costs. Publishability does not enter directly into student payoff, as short-term course outcomes typically determine student welfare more strongly than long-term dissemination benefits. Students therefore benefit indirectly from publishability only through their contribution to grade-relevant outcomes such as $Q$ and $D_{\text{doc}}$. Any long-run benefits of publishability to students are assumed to be second-order relative to course grades and are therefore omitted in this reduced form specification.

\end{itemize}

\subsection{Information structure, strategies, and beliefs}
\label{struture}
Because each player observes only part of the environment, beliefs about the types of the other players shape strategic behavior. The game proceeds sequentially. The university moves first, the sponsor moves second after observing the university’s policy, and the student team moves last after observing the actions of both the university and the sponsor. The information structure for the game is:  

\begin{enumerate}
    \item The university observes only its own type $\theta_U$.
    \item The sponsor observes its type $\theta_C$ and the university’s policy $(r,i,m)$.
    \item The student team observes its type $\theta_S$, the university’s policy, and the sponsor’s observable choices $(s,o,d)$.
\end{enumerate}

\noindent These information structures feed into the strategy functions. A pure strategy for each player maps its information set to an action
\begin{align}
    \sigma_U &: \Theta_U \to [0,1] \times \{\text{permissive}, \text{restrictive}\} \times \{0,1,2\} \\
    \sigma_C &: \Theta_C \times A_U \to \{\text{supportive},\text{exploitative}\} \times \{0,1,2\} \times [0,\bar{d}] \\
    \sigma_S &: \Theta_S \times A_U \times \mathcal{A}_C \to [0,1] \times \{\text{L},\text{D},\text{M}\}.
\end{align}
Let $A_U = [0,1] \times \{\text{permissive},\text{restrictive}\} \times \{0,1,2\}$ and $A_C = \{\text{supportive},\text{exploitative}\} \times \{0,1,2\} \times [0,\bar d]$ denote the university and sponsor action spaces, respectively. Because university policies are chosen strategically, observing $(r,i,m)$ allows the sponsor to update beliefs about $\theta_U$ through Bayes' rule, even though $\theta_U$ does not directly constrain the policy space. These strategy functions encode the sequential structure of the game; sponsors condition their behavior on the university’s policy, and students condition their behavior on both the university’s and sponsor’s actions. Because players have incomplete information about one another’s types, they maintain beliefs that are updated whenever observed actions provide information about type. Let $\mu_C(\theta_U \mid r,i,m)$ denote the sponsor’s posterior belief about the university type after observing the university’s policy $a_U=(r,i,m)$, and let $\mu_S(\theta_C \mid r,i,m,s,o,d)$ denote the student team’s posterior belief about the sponsor’s type after observing the sponsor’s action $a_C=(s,o,d)$.

On the equilibrium path, beliefs are updated using Bayes’ rule
\begin{equation}
    \mu_C(\theta_U \mid a_U)
    = 
    \frac{
        p(\theta_U)\, \ell(a_U \mid \theta_U)
    }{
        \sum_{\theta_U'} p(\theta_U')\, \ell(a_U \mid \theta_U')
    },
\end{equation}
\begin{equation}
   \mu_S(\theta_C \mid a_U,a_C)
    =
    \frac{
    f_C(\theta_C)\,\ell(a_C \mid a_U,\theta_C)
    }{
    \int_{0}^{1} f_C(t)\,\ell(a_C \mid a_U,t)\,dt
    }.
\end{equation}
where $\ell(a \mid \theta)$ denotes the likelihood of observing action $a$ when the corresponding player has type~$\theta$, as induced by that player’s strategy. Off the equilibrium path, beliefs may be specified in any manner consistent with equilibrium refinements. These posterior beliefs determine each player’s expected utility and therefore shape their optimal actions. The solution concept used in this paper is Perfect Bayesian Equilibrium, which requires sequential rationality and
Bayesian updating along the equilibrium path.

In the illustrative case studies presented in Section 4, beliefs are held fixed at representative values to emphasize equilibrium structure rather than belief dynamics. Bayesian updating enters the model by shaping expected utilities and therefore strategy selection, not by altering the reduced-form outcome functions themselves. In particular, posterior beliefs affect which sponsor postures and student orientations are rational responses to observed actions, but once a representative equilibrium configuration is specified, the outcome calculations condition on those strategies directly. The role of Bayesian updating is therefore to justify why particular strategic regimes arise under incomplete information, rather than to generate within-case numerical belief trajectories.

\subsection{Equilibrium concept}
The appropriate solution concept for this sequential game with incomplete information is PBE. Each player’s strategy must maximize its expected payoff given its beliefs about the types of the others. Explicitly:
\begin{itemize}
    \item The university chooses $(r,i,m) = \sigma_U(\theta_U)$ to maximize
    \begin{equation}
        \mathbb{E}_{\theta_C,\theta_S}
        \left[ 
            U_U(r,i,m; \sigma_C, \sigma_S)
        \right].
    \end{equation}
    \item The sponsor chooses $(s,o,d) = \sigma_C(\theta_C, r,i,m)$ to maximize
    \begin{equation}
        \mathbb{E}_{\theta_U,\theta_S}
        \left[
            U_C(s,o,d; r,i,m,\sigma_S)
            \; \middle|\;
            \mu_C(\theta_U \mid r,i,m)
        \right].
    \end{equation}
    \item The student team chooses $(e,x)=\sigma_S(\theta_S,r,i,m,s,o,d)$ to maximize
    \begin{equation}
        \mathbb{E}_{\theta_C}
        \left[
            U_S(e,x; r,i,m,s,o,d)
            \; \middle|\;
            \mu_S(\theta_C \mid r,i,m,s,o,d)
        \right].
    \end{equation}
\end{itemize}

\noindent On the equilibrium path, beliefs are updated according to Bayes’ rule
\begin{equation}
    \mu_C(\theta_U \mid r,i,m)
    = 
    \frac{p(\theta_U)\, \ell(r,i,m \mid \theta_U)}
         {\sum_{\theta_U'} p(\theta_U')\, \ell(r,i,m \mid \theta_U')},
\end{equation}
\begin{equation}
    \mu_S(\theta_C \mid r,i,m,s,o,d)
=
\frac{
f_C(\theta_C)\,\ell(s,o,d \mid r,i,m,\theta_C)
}{
\int_0^1 f_C(t)\,\ell(s,o,d \mid r,i,m,t)\,dt
}.
\end{equation}
where $\ell(a \mid \theta)$ is the likelihood of action $a$ under strategy $\sigma$ for type $\theta$. Off the equilibrium path, beliefs may be assigned arbitrarily, provided that they are compatible with equilibrium refinements. In this game, a profile
\begin{equation}
(\sigma_U, \sigma_C, \sigma_S, \mu_C, \mu_S)
\end{equation}
is a PBE if:
\begin{enumerate}
    \item Each player’s strategy is sequentially rational at every information set.
    \item On-path beliefs satisfy Bayes’ rule based on the equilibrium strategies.
    \item Off-path beliefs are specified in a way consistent with the structure of the game.
\end{enumerate}
This equilibrium concept ensures internal consistency between strategies, beliefs, and the information structure of the capstone project environment.

\subsection{Representative equilibrium regimes}
Depending on parameter configurations, the game can have several qualitatively distinct representative equilibrium regimes. These regimes summarize stable patterns of behavior in which the actions of each player constitute a best response to the expected actions of the others. These regimes should be interpreted as empirically recognizable equilibrium patterns rather than claims of uniqueness under all parameterizations. The three most obvious ones are: 

\begin{itemize}
\item \textbf{Cooperative educational equilibrium}: A cooperative educational equilibrium emerges when sponsor types are supportive, university policies are balanced, mentoring intensity is high, and students face incentives that reward learning and steady effort. In this regime, students optimally choose high effort $e$ and a learning-oriented strategy $x = L$, because supportive sponsors and moderate scope strictness $d$ make the investment in skill acquisition worthwhile. The sponsor provides a high intensity of mentoring $o=2$, and the university selects the strictness of the rubric $r$ that rewards both technical depth and documentation. The resulting project has high technical quality, strong documentation, and high publishability. This equilibrium reflects the intended pedagogical design of capstone programs: aligned incentives generate strong joint performance. In summary, this equilibrium happens when all player incentives are balanced and focused on project quality. 
\item \textbf{Exploitative sponsor equilibrium}: An exploitative sponsor equilibrium arises when sponsor types are exploitative, mentoring intensity is low or zero, and the sponsor imposes strict scope requirements $d$ without providing adequate support. Under these conditions, students optimally shift effort towards meeting sponsor expectations by adopting a deliverable-focused strategy $x = D$ while reducing overall effort $e$ because the learning returns are low. The university’s policies exert limited influence unless the mentoring requirements $m$ are binding; otherwise, the exploitative behavior of the sponsor determines the strategic environment. Technical quality is moderate to low, documentation suffers, timeliness becomes fragile, and sponsor alignment may appear high only because students prioritize superficial deliverable completion. This regime represents a misaligned equilibrium driven by sponsor overreach and insufficient institutional safeguards. This equilibrium happens when the sponsors expect significant engineering work to be done by the students at the expense of their educational experience and requirements. The focus is too much on project deliverables and too little on learning and academic work.  
\item \textbf{Grade-gaming equilibrium}: A grade-gaming equilibrium occurs when the university imposes very strict or documentation-heavy rubrics $r$, while sponsor engagement remains minimal and project scope is vague or loosely enforced. In this environment, students maximize the utility by selecting a minimalist or documentation-oriented strategy $x = M$ that satisfies the requirements of the rubric with minimal technical investment, directing the effort $e$ toward activities that improve their grades rather than learning or engineering results. Sponsors contribute little mentoring ($o=0$) and exert weak influence on deliverable expectations, making it rational for students to focus strategically on documentation artifacts rather than substantive engineering progress. The resulting project exhibits high documentation quality but weak technical depth, limited innovation, and low publishability. This equilibrium captures a common failure mode in capstone courses, where rubric incentives crowd out meaningful engineering work. This equilibrium happens when the focus on academic deliverables are excessive and discourages the students from doing engineering work in favor of academic exercises. The focus is too much on academic work (with the objectives of maximizing grades) and learning, and too little on the project deliverables to the sponsor. 
\end{itemize}


\subsection{Mechanism design problem for the university}
Because the university moves first in the sequential structure of the game, it effectively acts as a Stackelberg leader. By choosing observable policy variables $(r,i,m)$, the university anticipates how sponsors and student teams will respond in equilibrium. These responses are characterized by the PBE strategies
\begin{equation}
    \sigma_C^*(\theta_C; r,i,m), 
\end{equation}
\begin{equation}
    \sigma_S^*(\theta_S; r,i,m).
\end{equation}
For this game, let 
\begin{equation}
\sigma^*(\theta; r,i,m)
= 
\bigl(
    \sigma_U(\theta_U),
    \sigma_C^*(\theta_C; r,i,m),
    \sigma_S^*(\theta_S; r,i,m)
\bigr)
\end{equation}
denote the induced equilibrium strategy profile for type vector $\theta=(\theta_U,\theta_C,\theta_S)$ under the university's chosen policy. The expectation is taken over the joint distribution of player types, including heterogeneity in university priorities captured by $\theta_U$. Given policies $(r,i,m)$ and equilibrium strategies $\sigma^*$, the expected utilities of each stakeholder are
\begin{equation}
\mathbb{E}_\theta[U_U(\sigma^*)], 
\end{equation}
\begin{equation}
\mathbb{E}_\theta[U_S(\sigma^*)], 
\end{equation}
\begin{equation}
\mathbb{E}_\theta[U_C(\sigma^*)],
\end{equation}
where expectations are taken with respect to the common prior
$p(\theta) = p_U(\theta_U)f_C(\theta_C)p_S(\theta_S)$ where $f_C(\theta_C)$ denotes the probability density function of the continuous sponsor type on $[0,1]$. The university selects $(r,i,m)$ to maximize a weighted welfare objective
\begin{equation}
\max_{r,i,m}
\;
\mathbb{E}_\theta 
\left[
    U_U(\sigma^*) 
    + \lambda\, U_S(\sigma^*) 
    + \eta\, U_C(\sigma^*)
\right],
\end{equation}
where $\lambda$ and $\eta$ are scalarization weights expressing the university’s relative valuation of student welfare and sponsor satisfaction. Different choices of $(\lambda,\eta)$ trace out the policy trade-offs between pedagogical outcomes, equity considerations, and external partnership stability. The university must choose $(r,i,m)$ such that sponsor and student responses are
consistent with PBE strategies
\begin{align}
\sigma_C^*(\theta_C; r,i,m)
&\in 
\arg\max_{(s,o,d)}
\mathbb{E}_{\theta_U,\theta_S}
\bigl[
    U_C(s,o,d; r,i,m,\sigma_S^*)
    \mid \mu_C
\bigr],
\\
\sigma_S^*(\theta_S; r,i,m)
&\in
\arg\max_{(e,x)}
\mathbb{E}_{\theta_U,\theta_C}
\bigl[
    U_S(e,x; r,i,m,\sigma_C^*)
    \mid \mu_S
\bigr],
\end{align}
with beliefs $\mu_C$ and $\mu_S$ formed according to Bayes’ rule as described in Section~\ref{struture}. The policy choices must satisfy administrative feasibility $0 \leq r \leq 1, i \in \{\text{permissive},\text{restrictive}\}, m \in \{0,1,2\}$. This Stackelberg mechanism-design formulation captures how university policy shapes downstream behavior. By moving first and anticipating equilibrium responses, the university internalizes the strategic reactions of sponsors and students when determining optimal academic policies. The IC constraints ensure that sponsor and student actions are equilibrium best responses to the chosen policies, making the resulting equilibrium behavior consistent with the PBE analyzed previously.

\subsection{Justification for reduced form outcome functions}
The outcome functions used in this model are specified in reduced form affine representations. This choice follows standard practice in mechanism design and applied Bayesian games~\cite{bolton2005}, where the objective is to capture directional incentive effects rather than estimate structural production technologies. Linear representations ensure that marginal effects are transparent, parameters remain interpretable, and comparative statics can be derived without introducing inessential functional complexity. More elaborate nonlinear specifications could be introduced without changing the qualitative insights of the model but would obscure the connection between policy choices and stakeholder incentives. The reduced form approach therefore provides a tractable and analytically clear mapping from strategic choices to project outcomes while preserving the salient behavioral relationships observed in capstone collaborations. The purpose of these reduced form specifications is to make incentive effects and decision dependencies transparent, rather than to represent calibrated production relationships or to enable outcome prediction for specific programs.

Although the case studies presented later rely on representative parameter values, the qualitative result that sponsors possess limited unilateral leverage is structurally robust. Sponsor influence operates primarily through mentoring intensity and scope strictness, both of which enter outcome functions with diminishing or offsetting returns due to effort responses and cost terms. Varying payoff weights or outcome coefficients changes the magnitude of sponsor utility but does not eliminate the trade-off between extraction and educational quality. In this sense, the observed sponsor limitations reflect the structure of the incentive system rather than a particular parameter choice.

\subsection{Existence of equilibria}
Because each player’s action space is either finite (for posture, orientation, and mentoring-tier choices) or a compact interval of real numbers (for effort, rubric strictness, and scope strictness), and because all payoff and outcome functions are continuous in the players’ actions and types, at least one Perfect Bayesian Equilibrium (possibly in mixed or behavioral strategies) exists for this game. This follows from standard existence results for sequential games with incomplete information and compact strategy sets~\cite{krepswilson1982, bergin1989, gershkovmoldovanu2010}. The analysis in the remainder of the paper focuses on characterizing representative equilibrium regimes under different parameterizations of the reduced form model.

\subsection{Summary of the formal model}
The model integrates type uncertainty, sequential decision-making, and policy-dependent incentives in a capstone project environment that involves a university, an industry sponsor, and a student team. Each stakeholder possesses a privately known type that influences its incentives, observes a subset of the preceding actions, and forms beliefs using the Bayes’ rule where possible. The university first moves by choosing the strictness of the rubric, the intellectual property policy, and the minimum mentoring requirements. The sponsor then chooses its behavior posture, the intensity of the mentoring, and the strictness of the scope after observing the university’s policy. Finally, the student team selects the effort and orientation of the project after observing both previous stages. Reduced form outcome functions translate these choices into technical quality, documentation quality, timeliness, sponsor alignment, and publishability. Payoff functions combine these outcomes with effort costs and behavioral penalties, and the resulting strategic interaction is analyzed under the Perfect Bayesian Equilibrium refinement. The model provides a coherent foundation for exploring how policy, sponsor behavior, and student effort interact to shape the outcomes of the capstone project.

\section{Case Studies}
\label{Sec4}
\subsection{Case study 1: Highly collaborative capstone partnership}
Case Study 1 describes a capstone project in which the incentives are reasonably well aligned across the university, sponsor, and student team, less as an idealized best case than as a useful reference point for what has to go right for a project to run as intended. A sponsor approaches the capstone course with a genuinely supportive posture, checks in regularly, and provides meaningful technical guidance. In the model, this corresponds to a supportive stance with moderate mentoring intensity ($o = 1$) and a reasonably scoped project ($d = 0.3$). The university plays its part by setting a rubric of moderate strictness ($r = 0.6$), requiring a minimum level of mentoring engagement ($m = 1$), and allowing a permissive intellectual property policy ($i = \text{permissive}$) that enables students to share and publish aspects of their work. These choices reflect a balanced educational philosophy: Maintaining academic rigor while giving students sufficient flexibility to explore technical ideas and produce publicly valuable outcomes. The student team in this scenario behaves much like a well-functioning capstone group. Students invest substantial effort in the project ($e = 0.85$) and adopt a learning-oriented mindset ($x = L$), prioritizing conceptual understanding and skill development rather than focusing exclusively on deliverable completion. The sponsor’s type is modeled as academically supportive ($\theta_C = 0.8$), consistent with an industry partner motivated by long-term collaboration, educational impact, and sustained engagement with the program. These choices create a highly cooperative environment in which incentives are largely aligned across the university, sponsor, and students, while still reflecting the trade-offs inherent in a learning-oriented project. The parameter values used in this case are selected to produce outcomes strictly within the unit interval and to preserve the directional effects encoded in Equations~\ref{8}-\ref{12}, enabling a clear sensitivity analysis of how the model behaves under highly collaborative conditions.

Table~\ref{Tab1Parameters} summarizes the parameter and action configuration used for this scenario, while Table~\ref{Tab1Outcomes} reports the resulting project outcomes generated by the model. Technical quality is high ($Q \approx 0.863$), documentation quality is strong ($D_{\text{doc}} \approx 0.873$), and timeliness is excellent ($T \approx 0.805$), reflecting the combined effects of high student effort and sustained mentoring support. Sponsor alignment is moderately high ($A \approx 0.570$), capturing the trade-off between strong educational orientation and deliverable-driven alignment. Publishability is high ($P \approx 0.767$), driven by the permissive intellectual property policy, the sponsor’s academically supportive type, and the strong technical and documentation outcomes achieved under this configuration. Table~\ref{Tab1Utilities} reports the resulting stakeholder utilities. The students receive the largest payoff ($U_S \approx 0.721$), balancing learning gains against the disutility associated with high effort. The university also gets a high payoff ($U_U \approx 0.677$) benefiting from strong technical performance, documentation quality, and publishability. The sponsor achieves a moderate payoff ($U_C \approx 0.503$), reflecting favorable progress and alignment at an acceptable mentoring cost. The numerical values reported in Table~\ref{Tab1Parameters} are not empirical estimates but internally consistent parameters selected to illustrate the qualitative behavior of the model. The coefficients respect the sign structure specified in Equations~\ref{8}-\ref{12}, maintain normalized outcome values in $[0,1]$, and support a transparent sensitivity analysis of how equilibrium outcomes respond to changes in mentoring intensity, project scope, and institutional policy choices. This style of numerical illustration is standard in theoretical and mechanism-design modeling, where the objective is to clarify equilibrium behavior rather than to produce calibrated predictions. Alternative parameter values with the same sign structure yield qualitatively similar outcome regimes. In general, this case illustrates a highly cooperative outcome regime in which supportive sponsor behavior and balanced institutional policies yield mutually beneficial results for all stakeholders.

\begin{scriptsize}
\begin{longtable}{llllll}
\caption{Parameter and action configuration for Case Study 1 (cooperative educational equilibrium).}
\label{Tab1Parameters} \\
\hline
\textbf{Symbol} & \textbf{Value} & \textbf{Description} &
\textbf{Symbol} & \textbf{Value} & \textbf{Description} \\
\hline
\endfirsthead

\multicolumn{6}{c}{{\bfseries Table \thetable{} -- continued from previous page}} \\
\hline
\textbf{Symbol} & \textbf{Value} & \textbf{Description} &
\textbf{Symbol} & \textbf{Value} & \textbf{Description} \\
\hline
\endhead

\hline \multicolumn{6}{r}{{Continued on next page}} \\
\endfoot

\hline
\endlastfoot

\multicolumn{6}{l}{\textit{Outcome function coefficients}} \\

$q_0$ & 0.50 & Baseline technical quality &
$q_1$ & 0.30 & Effort effect on $Q$ \\

$q_2$ & 0.10 & Mentoring effect on $Q$ &
$q_3$ & 0.08 & Penalty for exploitative $s$ in $Q$ \\

$q_4$ & 0.04 & Scope penalty in $Q$ &
$q_5$ & 0.02 & Interaction effect in $Q$ \\

$d_0$ & 0.40 & Baseline documentation quality &
$d_1$ & 0.25 & Effort effect on $D_{\text{doc}}$ \\

$d_2$ & 0.20 & Bonus for $x=L$ in $D_{\text{doc}}$ &
$d_3$ & 0.05 & Rubric effect on $D_{\text{doc}}$ \\

$d_4$ & 0.03 & Mentoring effect on $D_{\text{doc}}$ &
$t_0$ & 0.50 & Baseline timeliness \\

$t_1$ & 0.20 & Effort effect on $T$ &
$t_2$ & 0.15 & Mentoring effect on $T$ \\

$t_3$ & 0.05 & Scope penalty in $T$ &
$a_0$ & 0.50 & Baseline alignment $A$ \\

$a_1$ & 0.30 & Bonus for $x=D$ in $A$ &
$a_2$ & 0.10 & Bonus for supportive $s$ in $A$ \\

$a_3$ & 0.10 & Scope penalty in $A$ &
$p_0$ & 0.10 & Baseline publishability \\

$p_1$ & 0.25 & Effect of permissive IP policy &
$p_2$ & 0.25 & Effect of sponsor type $\theta_C$ \\

$p_3$ & 0.15 & Quality effect on $P$ &
$p_4$ & 0.10 & Document effect on $P$ \\[4pt]

\multicolumn{6}{l}{\textit{Policy and action choices}} \\

$r$ & 0.60 & Rubric strictness &
$m$ & 1.00 & Mentoring requirement \\

$i$ & permissive & IP / publication policy &
$e$ & 0.85 & Student effort \\

$x$ & $L$ & Learning-focused orientation &
$s$ & supportive & Sponsor posture \\

$o$ & 1 & Mentoring intensity &
$d$ & 0.30 & Scope strictness \\

$\theta_C$ & 0.80 & Sponsor type &
 &  &  \\[4pt]

\multicolumn{6}{l}{\textit{Payoff weights and costs}} \\

$\alpha_Q$ & 0.40 & Weight on $Q$ in $U_U$ &
$\alpha_D$ & 0.30 & Weight on $D_{\text{doc}}$ in $U_U$ \\

$\alpha_P$ & 0.30 & Weight on $P$ in $U_U$ &
$c_r$ & 0.10 & Cost of rubric strictness \\

$c_m$ & 0.10 & Cost of mentoring requirement &
$\beta_A$ & 0.50 & Weight on $A$ in $U_C$ \\

$\beta_T$ & 0.40 & Weight on $T$ in $U_C$ &
$c_o$ & 0.10 & Cost of mentoring $o$ \\

$c_d$ & 0.05 & Cost of scope $d$ &
$\gamma_Q$ & 0.40 & Weight on $Q$ in $U_S$ \\

$\gamma_D$ & 0.30 & Weight on $D_{\text{doc}}$ in $U_S$ &
$\gamma_T$ & 0.30 & Weight on $T$ in $U_S$ \\

$k(\theta_S)$ & 0.15 & Effort cost for low effort-cost student types &
 &  &  \\

\end{longtable}
\end{scriptsize}

\begin{scriptsize}
\begin{longtable}{lll}
\caption{Model-generated outcomes for Case Study 1.}
\label{Tab1Outcomes} \\
\hline
\textbf{Outcome} & \textbf{Value} & \textbf{Interpretation} \\
\hline
\endfirsthead

\multicolumn{3}{c}%
{{\bfseries Table \thetable{} -- continued from previous page}} \\
\hline
\textbf{Outcome} & \textbf{Value} & \textbf{Interpretation} \\
\hline
\endhead

\hline
\endfoot

\hline
\endlastfoot

$Q$ & 0.863 & Technical quality \\
$D_{\text{doc}}$ & 0.873 & Documentation quality \\
$T$ & 0.805 & Timeliness of project completion \\
$A$ & 0.570 & Alignment with sponsor expectations \\
$P$ & 0.767 & Publishability of project outputs \\

\end{longtable}
\end{scriptsize}

\begin{scriptsize}
\begin{longtable}{lll}
\caption{Stakeholder utilities for Case Study 1.}
\label{Tab1Utilities} \\
\hline
\textbf{Stakeholder} & \textbf{Utility} & \textbf{Interpretation} \\
\hline
\endfirsthead

\multicolumn{3}{c}%
{{\bfseries Table \thetable{} -- continued from previous page}} \\
\hline
\textbf{Stakeholder} & \textbf{Utility} & \textbf{Interpretation} \\
\hline
\endhead

\hline
\endfoot

\hline
\endlastfoot

University ($U_U$) & 0.677 & Gains from learning, documentation, and publishability \\
Sponsor ($U_C$) & 0.503 & Benefits from alignment and timeliness net of mentoring costs \\
Students ($U_S$) & 0.721 & Gains from outcomes net of effort disutility \\

\end{longtable}
\end{scriptsize}

\textbf{Equilibrium verification}: Given the action profile specified in Case Study~1, the resulting configuration is locally best-response consistent for all three players under the reduced form payoff and outcome mappings. For the university, small deviations in rubric strictness $r$ or mentoring requirement $m$ do not yield higher expected utility at this operating point, as the induced changes in documentation quality and publishability are insufficient to generate a net improvement in $U_U$. For the sponsor, reducing mentoring intensity $o$ lowers timeliness and technical quality, while increasing project scope $d$ reduces alignment and timeliness, making either deviation unprofitable given the specified weights. For the student team, reducing effort $e$ decreases technical quality, documentation quality, and timeliness enough to offset the reduction in effort disutility, and deviating from the learning-oriented strategy $x=L$ removes learning-related contributions to project outcomes. Consequently, no player can improve its expected utility through a small unilateral deviation, and the Case Study~1 configuration constitutes a locally best-response consistent equilibrium instance of the reduced form Bayesian game.

\subsection{Case study 2: Exploitative sponsor equilibrium}
Case Study 2 examines a capstone project dominated by sponsor incentives, reflecting the familiar situation in which an external partner treats the project primarily as a mechanism for extracting deliverables rather than as an educational collaboration. In the model, this corresponds to a sponsor that adopts an explicitly exploitative posture ($s=\text{exploitative}$), provides no meaningful mentoring support ($o=0$), and imposes a highly demanding project scope ($d=0.8$). The sponsor’s underlying type reflects weak academic orientation ($\theta_C=0.2$), consistent with limited interest in learning outcomes or dissemination potential. At the outset of the project, the university does not observe the sponsor’s engagement intentions and therefore applies the same baseline policies used in the cooperative scenario. In particular, the university maintains a moderately strict grading rubric ($r=0.6$), requires a minimum mentoring level ($m=1$), and adopts a permissive intellectual property policy ($i=\text{permissive}$), reflecting institutional priorities that favor academic rigor and the possibility of publishable outcomes. However, because the mentoring requirement is not directly enforceable during the semester, these policies do not induce increased sponsor engagement. The absence of mentoring combined with the aggressive project scope quickly constrains the learning environment faced by the student team.

In response to high expectations, limited feedback, and tight timelines, the students adjust their strategy in a manner consistent with rational effort allocation. In the model, the team reduces overall effort to a moderate level ($e=0.40$) and shifts away from a learning-oriented approach toward a deliverable-focused orientation ($x=D$). Rather than investing in exploration or skill development, students prioritize producing minimally acceptable outputs under time pressure, a behavior commonly observed in poorly supported industry-sponsored projects. The resulting outcomes, summarized in Tables~\ref{Tab2Parameters}-\ref{Tab2Utilities}, reflect this misalignment of incentives. Technical quality, documentation quality, and timeliness are all substantially lower than in the cooperative case, while sponsor alignment appears relatively high due to the students’ emphasis on deliverable completion. Publishability is correspondingly reduced, driven by weaker technical outcomes and limited documentation depth. Although university policies remain unchanged from Case Study~1, the shift in sponsor behavior alone is sufficient to push the system into a low-learning, compliance-driven regime. This case study therefore illustrates how exploitative sponsor incentives can dominate the strategic environment and degrade educational outcomes even under otherwise well-designed institutional policies.

\begin{scriptsize}
\begin{longtable}{llllll}
\caption{Parameter and action configuration for Case Study 2 (exploitative sponsor equilibrium).}
\label{Tab2Parameters} \\
\hline
\textbf{Symbol} & \textbf{Value} & \textbf{Description} &
\textbf{Symbol} & \textbf{Value} & \textbf{Description} \\
\hline
\endfirsthead

\multicolumn{6}{c}{{\bfseries Table \thetable{} -- continued from previous page}} \\
\hline
\textbf{Symbol} & \textbf{Value} & \textbf{Description} &
\textbf{Symbol} & \textbf{Value} & \textbf{Description} \\
\hline
\endhead

\hline \multicolumn{6}{r}{{Continued on next page}} \\
\endfoot

\hline
\endlastfoot

\multicolumn{6}{l}{\textit{Outcome function coefficients}} \\

$q_0$ & 0.50 & Baseline technical quality &
$q_1$ & 0.30 & Effort effect on $Q$ \\

$q_2$ & 0.10 & Mentoring effect on $Q$ &
$q_3$ & 0.08 & Penalty for exploitative $s$ in $Q$ \\

$q_4$ & 0.04 & Scope penalty in $Q$ &
$q_5$ & 0.02 & Interaction effect in $Q$ \\

$d_0$ & 0.40 & Baseline documentation quality &
$d_1$ & 0.25 & Effort effect on $D_{\text{doc}}$ \\

$d_2$ & 0.20 & Bonus for $x=L$ in $D_{\text{doc}}$ &
$d_3$ & 0.05 & Rubric effect on $D_{\text{doc}}$ \\

$d_4$ & 0.03 & Mentoring effect on $D_{\text{doc}}$ &
$t_0$ & 0.50 & Baseline timeliness \\

$t_1$ & 0.20 & Effort effect on $T$ &
$t_2$ & 0.15 & Mentoring effect on $T$ \\

$t_3$ & 0.05 & Scope penalty in $T$ &
$a_0$ & 0.50 & Baseline alignment $A$ \\

$a_1$ & 0.30 & Bonus for $x=D$ in $A$ &
$a_2$ & 0.10 & Bonus for supportive $s$ in $A$ \\

$a_3$ & 0.10 & Scope penalty in $A$ &
$p_0$ & 0.10 & Baseline publishability \\

$p_1$ & 0.25 & Effect of permissive IP policy &
$p_2$ & 0.25 & Effect of sponsor type $\theta_C$ \\

$p_3$ & 0.15 & Quality effect on $P$ &
$p_4$ & 0.10 & Document effect on $P$ \\[4pt]

\multicolumn{6}{l}{\textit{Policy and action choices}} \\

$r$ & 0.60 & Rubric strictness &
$m$ & 1.00 & Mentoring requirement \\

$i$ & permissive & IP / publication policy &
$e$ & 0.40 & Student effort \\

$x$ & $D$ & Deliverable-focused orientation &
$s$ & exploitative & Sponsor posture \\

$o$ & 0 & Mentoring intensity &
$d$ & 0.80 & Scope strictness \\

$\theta_C$ & 0.20 & Sponsor type &
 &  &  \\[4pt]

\multicolumn{6}{l}{\textit{Payoff weights and costs}} \\

$\alpha_Q$ & 0.40 & Weight on $Q$ in $U_U$ &
$\alpha_D$ & 0.30 & Weight on $D_{\text{doc}}$ in $U_U$ \\

$\alpha_P$ & 0.30 & Weight on $P$ in $U_U$ &
$c_r$ & 0.10 & Cost of rubric strictness \\

$c_m$ & 0.10 & Cost of mentoring requirement &
$\beta_A$ & 0.50 & Weight on $A$ in $U_C$ \\

$\beta_T$ & 0.40 & Weight on $T$ in $U_C$ &
$c_o$ & 0.10 & Cost of mentoring $o$ \\

$c_d$ & 0.05 & Cost of scope $d$ &
$\gamma_Q$ & 0.40 & Weight on $Q$ in $U_S$ \\

$\gamma_D$ & 0.30 & Weight on $D_{\text{doc}}$ in $U_S$ &
$\gamma_T$ & 0.30 & Weight on $T$ in $U_S$ \\

$k(\theta_S)$ & 0.15 & Effort cost for low effort-cost student types &
 &  &  \\

\end{longtable}
\end{scriptsize}

\begin{scriptsize}
\begin{longtable}{lll}
\caption{Model-generated outcomes for Case Study 2.}
\label{Tab2Outcomes} \\
\hline
\textbf{Outcome} & \textbf{Value} & \textbf{Interpretation} \\
\hline
\endfirsthead

\multicolumn{3}{c}%
{{\bfseries Table \thetable{} -- continued from previous page}} \\
\hline
\textbf{Outcome} & \textbf{Value} & \textbf{Interpretation} \\
\hline
\endhead

\hline
\endfoot

\hline
\endlastfoot

$Q$ & 0.508 & Technical quality \\
$D_{\text{doc}}$ & 0.530 & Documentation quality \\
$T$ & 0.540 & Timeliness of project completion \\
$A$ & 0.720 & Alignment with sponsor expectations \\
$P$ & 0.529 & Publishability of project outputs \\

\end{longtable}
\end{scriptsize}

\begin{scriptsize}
\begin{longtable}{lll}
\caption{Stakeholder utilities for Case Study 2.}
\label{Tab2Utilities} \\
\hline
\textbf{Stakeholder} & \textbf{Utility} & \textbf{Interpretation} \\
\hline
\endfirsthead

\multicolumn{3}{c}%
{{\bfseries Table \thetable{} -- continued from previous page}} \\
\hline
\textbf{Stakeholder} & \textbf{Utility} & \textbf{Interpretation} \\
\hline
\endhead

\hline
\endfoot

\hline
\endlastfoot

University ($U_U$) & 0.361 & Gains from learning, documentation, and publishability \\
Sponsor ($U_C$) & 0.544 & Benefits from alignment and timeliness net of mentoring costs \\
Students ($U_S$) & 0.464 & Gains from outcomes net of effort disutility \\

\end{longtable}
\end{scriptsize}

\textbf{Equilibrium verification}: Given the action profile specified in Case Study~2, the resulting configuration is locally best-response consistent for all three players under the reduced form payoff and outcome mappings. For the sponsor, adopting an exploitative posture $s=\text{exploitative}$ together with zero mentoring $o=0$ is locally optimal at this operating point: Increasing $o$ raises mentoring costs while yielding only limited marginal gains in alignment and timeliness under the induced student response. Similarly, tightening scope beyond the chosen level $d=0.8$ further reduces alignment and timeliness and increases the scope-related cost term, so that small deviations in $d$ do not improve $U_C$. Taking the sponsor's behavior as given, the student team responds by selecting a deliverable-focused orientation $x=D$ and moderate effort $e=0.40$. Local increases in effort raise the disutility term $k(\theta_S)e$ while producing only modest improvements in technical quality, documentation quality, and timeliness when mentoring is absent. Deviating toward a learning-oriented strategy $x=L$ removes the deliverable-alignment advantage without generating compensating gains through mentoring-driven interaction effects, and therefore does not increase $U_S$ at this point. Finally, given the sponsor and student responses, small deviations in the university policy variables $(r,i,m)$ do not yield higher expected utility at this operating point. Consequently, no player can improve its expected utility through a small unilateral deviation, and the Case Study~2 configuration constitutes a locally best-response consistent equilibrium instance of the reduced form Bayesian game representing an exploitative sponsor regime.

\subsection{Case study 3: Grade-gaming equilibrium}
Case Study 3 considers a capstone project shaped largely by assessment incentives, capturing a common pattern in which students respond rationally to grading structures by optimizing for rubric compliance and assignment grades rather than for technical depth or exploration. Such scenarios are common when grading rubrics place heavy emphasis on documentation, formal deliverables, and compliance with prescribed requirements, thereby signaling to students that grades depend more on rubric-visible outputs than on substantive technical depth or exploratory learning. In the model, this corresponds to a university that adopts a highly stringent grading rubric ($r=0.9$) while maintaining a permissive intellectual property policy ($i=\text{permissive}$) and a moderate mentoring expectation ($m=1$). The sponsor in this scenario adopts a supportive posture ($s=\text{supportive}$) and provides a moderate level of mentoring ($o=1$), consistent with periodic design reviews and progress check-ins rather than intensive technical involvement. The sponsor’s underlying type is reasonably aligned with academic objectives ($\theta_C=0.6$), and the project scope is challenging but not extreme ($d=0.6$). Despite the sponsor’s willingness to engage, the university’s grading structure dominates the incentive landscape faced by the student team.

In response to these assessment signals, students adjust their behavior to prioritize activities that are most directly rewarded by the grading rubric. In the model, the team allocates a moderate level of overall effort ($e=0.60$) but adopts a minimalist technical orientation ($x=M$), avoiding exploratory or high-risk engineering work that may not translate into immediate rubric points. Instead, effort is concentrated on producing well-formatted documentation, meeting formal requirements, and managing deadlines. This response reflects rational grade-maximizing behavior rather than a lack of motivation or capability. The resulting outcomes, summarized in Tables~\ref{Tab3Parameters}-\ref{Tab3Utilities}, reflect this shift in strategic emphasis. Documentation quality and timeliness remain relatively strong, while technical quality and publishability are reduced compared to the cooperative case. Sponsor alignment remains moderate, as required deliverables are produced on schedule, but the sponsor derives limited value from the project’s technical depth. Although sponsor behavior is not exploitative in this scenario, the university’s emphasis on assessment rigor alone is sufficient to induce a grade-gaming regime in which compliance crowds out deeper engineering learning. This case study illustrates how well-intentioned academic policies can inadvertently drive strategic student behavior toward locally optimal but educationally shallow outcomes.

\begin{scriptsize}
\begin{longtable}{llllll}
\caption{Parameter and action configuration for Case Study 3 (grade-gaming equilibrium).}
\label{Tab3Parameters} \\
\hline
\textbf{Symbol} & \textbf{Value} & \textbf{Description} &
\textbf{Symbol} & \textbf{Value} & \textbf{Description} \\
\hline
\endfirsthead

\multicolumn{6}{c}{{\bfseries Table \thetable{} -- continued from previous page}} \\
\hline
\textbf{Symbol} & \textbf{Value} & \textbf{Description} &
\textbf{Symbol} & \textbf{Value} & \textbf{Description} \\
\hline
\endhead

\hline \multicolumn{6}{r}{{Continued on next page}} \\
\endfoot

\hline
\endlastfoot

\multicolumn{6}{l}{\textit{Outcome function coefficients}} \\

$q_0$ & 0.50 & Baseline technical quality &
$q_1$ & 0.30 & Effort effect on $Q$ \\

$q_2$ & 0.10 & Mentoring effect on $Q$ &
$q_3$ & 0.08 & Penalty for exploitative $s$ in $Q$ \\

$q_4$ & 0.04 & Scope penalty in $Q$ &
$q_5$ & 0.02 & Interaction effect in $Q$ \\

$d_0$ & 0.40 & Baseline documentation quality &
$d_1$ & 0.25 & Effort effect on $D_{\text{doc}}$ \\

$d_2$ & 0.20 & Bonus for $x=L$ in $D_{\text{doc}}$ &
$d_3$ & 0.05 & Rubric effect on $D_{\text{doc}}$ \\

$d_4$ & 0.03 & Mentoring effect on $D_{\text{doc}}$ &
$t_0$ & 0.50 & Baseline timeliness \\

$t_1$ & 0.20 & Effort effect on $T$ &
$t_2$ & 0.15 & Mentoring effect on $T$ \\

$t_3$ & 0.05 & Scope penalty in $T$ &
$a_0$ & 0.50 & Baseline alignment $A$ \\

$a_1$ & 0.30 & Bonus for $x=D$ in $A$ &
$a_2$ & 0.10 & Bonus for supportive $s$ in $A$ \\

$a_3$ & 0.10 & Scope penalty in $A$ &
$p_0$ & 0.10 & Baseline publishability \\

$p_1$ & 0.25 & Effect of permissive IP policy &
$p_2$ & 0.25 & Effect of sponsor type $\theta_C$ \\

$p_3$ & 0.15 & Quality effect on $P$ &
$p_4$ & 0.10 & Document effect on $P$ \\[4pt]

\multicolumn{6}{l}{\textit{Policy and action choices}} \\

$r$ & 0.90 & Rubric strictness &
$m$ & 1.00 & Mentoring requirement \\

$i$ & permissive & IP / publication policy &
$e$ & 0.60 & Student effort \\

$x$ & $M$ & Minimalist-focused orientation &
$s$ & supportive & Sponsor posture \\

$o$ & 1 & Mentoring intensity &
$d$ & 0.60 & Scope strictness \\

$\theta_C$ & 0.60 & Sponsor type &
 &  &  \\[4pt]

\multicolumn{6}{l}{\textit{Payoff weights and costs}} \\

$\alpha_Q$ & 0.40 & Weight on $Q$ in $U_U$ &
$\alpha_D$ & 0.30 & Weight on $D_{\text{doc}}$ in $U_U$ \\

$\alpha_P$ & 0.30 & Weight on $P$ in $U_U$ &
$c_r$ & 0.10 & Cost of rubric strictness \\

$c_m$ & 0.10 & Cost of mentoring requirement &
$\beta_A$ & 0.50 & Weight on $A$ in $U_C$ \\

$\beta_T$ & 0.40 & Weight on $T$ in $U_C$ &
$c_o$ & 0.10 & Cost of mentoring $o$ \\

$c_d$ & 0.05 & Cost of scope $d$ &
$\gamma_Q$ & 0.40 & Weight on $Q$ in $U_S$ \\

$\gamma_D$ & 0.30 & Weight on $D_{\text{doc}}$ in $U_S$ &
$\gamma_T$ & 0.30 & Weight on $T$ in $U_S$ \\

$k(\theta_S)$ & 0.15 & Effort cost for low effort-cost student types &
 &  &  \\

\end{longtable}
\end{scriptsize}

\begin{scriptsize}
\begin{longtable}{lll}
\caption{Model-generated outcomes for Case Study 3.}
\label{Tab3Outcomes} \\
\hline
\textbf{Outcome} & \textbf{Value} & \textbf{Interpretation} \\
\hline
\endfirsthead

\multicolumn{3}{c}%
{{\bfseries Table \thetable{} -- continued from previous page}} \\
\hline
\textbf{Outcome} & \textbf{Value} & \textbf{Interpretation} \\
\hline
\endhead

\hline
\endfoot

\hline
\endlastfoot

$Q$ & 0.756 & Technical quality \\
$D_{\text{doc}}$ & 0.625 & Documentation quality \\
$T$ & 0.740 & Timeliness of project completion \\
$A$ & 0.540 & Alignment with sponsor expectations \\
$P$ & 0.676 & Publishability of project outputs \\

\end{longtable}
\end{scriptsize}

\begin{scriptsize}
\begin{longtable}{lll}
\caption{Stakeholder utilities for Case Study 3.}
\label{Tab3Utilities} \\
\hline
\textbf{Stakeholder} & \textbf{Utility} & \textbf{Interpretation} \\
\hline
\endfirsthead

\multicolumn{3}{c}%
{{\bfseries Table \thetable{} -- continued from previous page}} \\
\hline
\textbf{Stakeholder} & \textbf{Utility} & \textbf{Interpretation} \\
\hline
\endhead

\hline
\endfoot

\hline
\endlastfoot

University ($U_U$) & 0.503 & Gains from learning, documentation, and publishability \\
Sponsor ($U_C$) & 0.448 & Benefits from alignment and timeliness net of mentoring costs \\
Students ($U_S$) & 0.622 & Gains from outcomes net of effort disutility \\

\end{longtable}
\end{scriptsize}

\textbf{Equilibrium verification}: Given the action profile specified in Case Study~3, the resulting configuration is locally best-response consistent for all three players under the reduced form payoff and outcome mappings. For the student team, the combination of a highly stringent grading rubric ($r=0.9$) and moderate mentoring support induces a grading-dominant incentive environment. At this operating point, adopting a minimalist technical orientation $x=M$ together with moderate effort $e=0.60$ maximizes student utility. Deviating toward a learning-oriented strategy $x=L$ reduces documentation-related contributions to outcomes that are heavily weighted by the rubric, while the incremental gains in technical quality are insufficient to offset this loss. Similarly, increasing effort beyond the chosen level raises effort disutility without generating proportional improvements in rubric-visible outcomes. Given the induced student response, the sponsor’s choice of a supportive posture with moderate mentoring intensity ($s=\text{supportive}$, $o=1$) is locally optimal. Increasing mentoring intensity would raise sponsor costs while yielding limited marginal gains in alignment or project value, as students are not pursuing deeper technical exploration. Reducing mentoring, on the other hand, would lower timeliness and technical quality without improving sponsor utility at this point. Finally, taking sponsor and student behavior as given, the university cannot improve expected utility through small deviations in policy variables. Relaxing rubric strictness would directly reduce documentation quality and publishability, while further tightening requirements does not induce additional student effort or learning-oriented behavior. Adjustments to mentoring expectations likewise do not produce utility gains given the sponsor’s and students’ equilibrium responses. Consequently, no player can improve its expected utility through a small unilateral deviation, and the Case Study~3 configuration constitutes a locally best-response consistent equilibrium instance of the reduced form Bayesian game, representing a grade-gaming regime driven by assessment incentives.

\begin{figure}[h!]
\centering
\includegraphics[width=1\textwidth]{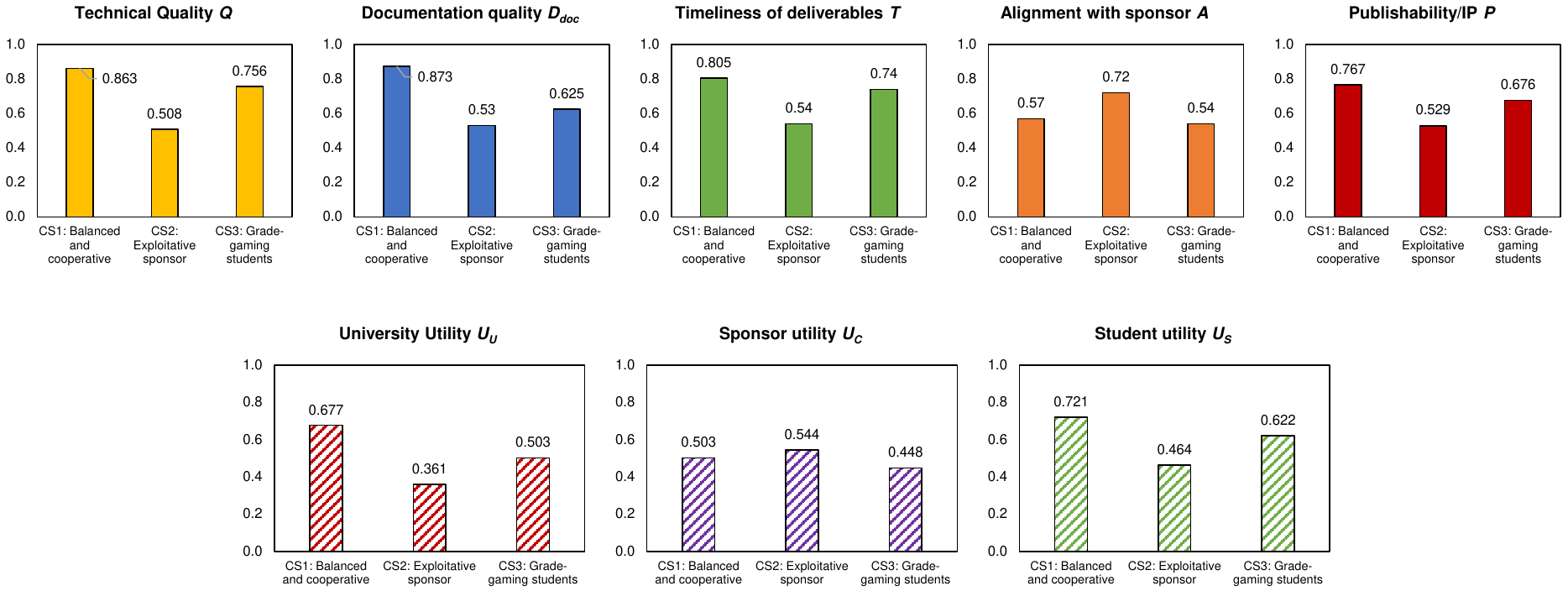}
\caption{Stakeholder utilities across equilibrium regimes, illustrating incentive dominance. Sponsor utility is maximized in CS2, student utility in CS3, and balanced returns emerge only under cooperative incentives (CS1).}
\label{plots}
\end{figure}

\subsection{Cross-Case Synthesis of Outcomes and Utilities}
To facilitate comparison between the three representative equilibrium regimes, Figures~\ref{plots} summarize the outcomes achieved and the utilities of stakeholders, respectively. The purpose of this synthesis is not to introduce new results, but to highlight cross-case patterns that are difficult to infer directly from the individual case tables. Figure~\ref{plots} illustrates how different incentive structures produce qualitatively different outcome profiles despite fixed model coefficients. In the cooperative regime (CS1), all outcome metrics achieve relatively high values, reflecting balanced incentives and mutual engagement among stakeholders. In contrast, the exploitative sponsor regime (CS2) exhibits reduced technical quality and documentation quality despite strong sponsor alignment, indicating that deliverable extraction can coexist with degraded learning and communication outcomes. The grade-gaming regime (CS3) shows a complementary pattern: Documentation quality remains high under strict assessment, while technical quality and publishability decline, reflecting compliance with grading incentives rather than deep technical engagement. Figure~\ref{plots} further clarifies the incentive dominance underlying these regimes. In CS1, stakeholder utilities are comparatively well aligned, supporting sustained cooperation. In CS2, the sponsor utility dominates, while the student and university utilities are comparatively lower, capturing a regime in which value extraction is prioritized over educational outcomes. In CS3, student utility is comparatively strong despite reduced sponsor and university returns, consistent with rational grade--maximizing behavior under stringent evaluation criteria. Taken together, these visual summaries reinforce a central insight of the model: Capstone outcomes are governed less by individual goodwill than by the incentives of the stakeholders that dominate the strategic environment. Therefore, similar institutional policies can give rise to markedly different educational and technical outcomes, depending on how grading structures, mentoring behavior, and sponsor engagement interact to shape equilibrium behavior. This synthesis motivates the broader discussion in Section~\ref{Sec5} on incentive dominance, policy tradeoffs, and structural failure modes in industry--sponsored capstone design.

\section{Discussion and Analysis}
\label{Sec5}
This paper develops a reduced form game-theoretic framework for analyzing industry-sponsored capstone projects as strategic systems governed by interacting incentives rather than as isolated educational experiences. By explicitly modeling the university, sponsor, and student team as rational agents with distinct (and perhaps conflicting) objectives, the framework explains how stable outcome patterns emerge from within stakeholder interactions and team structure, even when all stakeholders act in good faith. The resulting equilibria closely correspond to well-documented capstone project regimes and provide insight into why certain pathologies persist despite careful program design.

\subsection{Rationality and equilibrium: Persistent problems}
A central contribution of the model is to demonstrate that undesirable capstone outcomes need not arise from negligence, miscommunication, or lack of commitment by any individual stakeholder. In all three case studies, the university, the sponsor, and the student team behave in a locally rational manner, selecting strategies that maximize the expected utility given the actions of the others. Importantly, the exploitative sponsor regime and the grade-gaming regime are not transient failures but stable equilibrium configurations in which no stakeholder can unilaterally improve outcomes through small deviations. This equilibrium perspective re-frames common capstone challenges. Rather than attributing low learning outcomes to unmotivated students or disengaged sponsors, the model shows that such behaviors are often optimal responses to the incentive environment. Once an equilibrium regime is established under a fixed policy and monitoring structure, appeals to professionalism, goodwill, or educational mission are unlikely to be effective unless the underlying incentives or information structure are altered. This insight helps explain why repeated attempts to correct capstone dysfunction through incremental policy adjustments often yield limited improvement.

This formulation intentionally abstracts away from mid-course policy revision and adaptive monitoring by the university. In practice, faculty advisors and program administrators may intervene by adjusting scope, increasing oversight, or renegotiating expectations in response to emerging issues. Such interventions can be interpreted as shifts in the information structure or as partial re-commitments to revised policies. Incorporating these dynamics would require a repeated-game or adaptive control formulation, which lies beyond the scope of the present analysis but represents a natural direction for future work.

\subsection{Dominant-incentive regimes and stakeholder control of outcomes}
In all three case studies, the results are governed by the incentives of the stakeholders that dominate the strategic environment. In the cooperative regime, the objectives of no one actor overwhelm the others, allowing learning-oriented behavior to emerge naturally from within the stakeholder team. In contrast, the exploitative sponsor regime and the grade-gaming regime illustrate how dominance by a single stakeholder collapses the system into a single-objective optimization problem. When sponsor incentives dominate, as in Case Study~2, student behavior shifts toward deliverable compliance and scope containment, while learning and exploration are crowded out. In contrast, when university assessment incentives dominate, as in the Case Study~3, students rationally prioritize rubric-visible outputs (related to grades) at the expense of technical exploration on the project itself. In particular, these regimes arise even when other stakeholders behave reasonably: A supportive sponsor cannot fully counteract a grading structure that strongly rewards compliance, and well-designed university policies cannot overcome sponsor disengagement when mentoring is unenforceable. This dominant-incentive perspective provides a unifying explanation for a wide range of observed capstone outcomes and highlights the fragility of balanced incentive structures. Small changes in sponsor behavior or assessment emphasis can shift the system into a qualitatively different equilibrium, suggesting that successful capstone programs operate near the boundary between regimes rather than deep within a stable optimum.

\subsection{Limits of policy-centric approaches to capstone design}
An important implication of the analysis is that strong institutional policies alone are insufficient to guarantee desirable educational outcomes. Across the case studies, university grading and intellectual property policies are held constant, yet outcomes vary dramatically as sponsor behavior and student strategies change. This shows that policy quality is a necessary but not sufficient condition for success. The model further suggests that certain policy levers have diminishing returns when deployed in isolation. Increasing rubric strictness may improve documentation quality but can inadvertently encourage grade gaming. Similarly, nominal mentoring requirements are ineffective when they cannot be enforced or when sponsors lack intrinsic motivation to engage. These findings challenge the assumption that capstone outcomes can be optimized through isolated policy refinement and instead point to the need for coordinated incentive alignment between stakeholders.

\subsection{Interpretation of outcome metrics and apparent success}
The case studies also reveal that commonly used performance metrics can be misleading indicators of educational success. High sponsor alignment, timely completion of the deliverable, or strong documentation quality can coexist with shallow learning and limited technical development. In the exploitative sponsor regime, alignment appears relatively high precisely because students prioritize sponsor-defined deliverables, even as technical quality and publishability decline. In the grade-gaming regime, documentation quality remains strong while exploratory learning is suppressed. These results underscore the importance of interpreting the outcomes of the capstone in light of the incentive structures that produce them. Metrics that are valuable for accountability or assessment may not reliably reflect educational depth unless they are embedded within a balanced incentive system. The framework therefore provides a lens for diagnosing when apparent success masks deeper structural issues.

\subsection{Equilibrium regimes as explanatory, not predictive, tools}
Each equilibrium regime should be interpreted as a diagnostic representation of which stakeholder incentives dominate the strategic environment, rather than as a retrospective label of success or failure. The equilibrium regimes presented in this work are also not intended as unique predictions of capstone outcomes under specific parameter values. Rather, they should be interpreted as representative configurations that capture empirically recognizable patterns of behavior. The reduced form structure of the model enables a clear identification of incentive-driven dynamics, but it necessarily abstracts away from contextual variation between institutions, disciplines, and project types. This distinction is important in avoiding over-interpretation. The value of the framework lies in its explanatory power, it clarifies why certain patterns recur across diverse settings and why interventions that ignore strategic interactions often fail. By focusing on equilibrium behavior, the model provides insight into structural forces that shape capstone experiences beyond any single implementation.

\subsection{Design implications and leverage points}
Although the paper does not prescribe specific program designs, the analysis suggests several leverage points to improve capstone outcomes. First, enforceability matters: Mentoring expectations and engagement requirements must be coupled to mechanisms that make deviation costly or visible. Second, assessment structures send powerful signals; grading rubrics that overemphasize compliance can unintentionally suppress learning even when other supports are present. Third, sponsor selection and screening may be more effective than downstream policy adjustments in preventing exploitative regimes. More broadly, the results indicate that capstone design should be approached as a mechanism design problem under institutional constraints. Effective interventions are those that shift the system away from dominant-incentive regimes and toward balanced configurations where learning-oriented behavior is individually rational for all stakeholders.

\subsection{Limitations and opportunities for extension}
The model intentionally adopts a reduced form, single period structure to isolate incentive interactions. It does not capture repeated interactions across cohorts, reputation effects between sponsors and institutions, spillovers across simultaneous project teams, or within team heterogeneity, all of which are important features of real capstone programs. Faculty advisors are treated implicitly as part of the institutional mechanism rather than as independent strategic actors with distinct incentives and constraints. These omissions are deliberate and allow the analysis to focus on equilibrium structure rather than on dynamic or program specific complexity. Future work could extend the framework to repeated game settings in which sponsors and universities learn from prior outcomes, or to models with heterogeneous student types and internal team dynamics. Incorporating sponsor screening, signaling, or contract mechanisms would further enrich the analysis and connect the framework to broader literature in mechanism design and organizational economics.

\subsection{Broader significance}
The central insight of this work is that \textit{capstone project outcomes are governed less by individual goodwill than by which stakeholder’s incentives dominate the strategic environment}. By modeling capstone projects as strategic systems, the framework provides a principled explanation for persistent educational challenges and a foundation for more effective design interventions. Recognizing and managing incentive dominance is therefore a necessary condition for creating capstone experiences that reliably support deep learning, meaningful industry engagement, and institutional objectives.

\section{Closing Remarks}
\label{Sec6}
This paper presents a formal game-theoretic framework for analyzing industry-sponsored engineering capstone projects as strategic systems shaped by interacting incentives rather than as isolated educational experiences. By modeling the university, sponsor, and student team as rational actors operating under incomplete information, the framework explains how familiar capstone outcomes emerge endogenously, including productive collaborations and persistent failure modes. Importantly, the analysis demonstrates that these outcomes can arise even when all stakeholders act in good faith and institutional policies remain unchanged. Case studies highlight that capstone success is not determined solely by policy quality, sponsor intent, or student motivation in isolation, but by which incentives dominate the strategic environment. When incentives are balanced, learning-oriented behavior can naturally emerge. When dominance shifts toward sponsor extraction or assessment compliance, rational responses from students and sponsors can crowd out deeper educational objectives. These dynamics help explain why incremental policy adjustments often fail to correct persistent capstone challenges. The framework is not intended to predict specific project outcomes or prescribe universal design solutions. Rather, it provides a principled lens for reasoning about incentive alignment, policy tradeoffs, and structural vulnerabilities in project-based learning environments. By making incentive interactions explicit, the model supports a more informed discussion among educators and program designers about where and why capstone systems succeed or fail. More broadly, this work positions the design of the capstone project as a problem of constrained mechanism design within engineering education. Future extensions incorporating repeated interactions, sponsor screening, or empirical calibration may further enrich the analysis. The central takeaway remains that long-term improvement in capstone outcomes requires attention not only to individual behavior but also to the strategic structures that govern how stakeholders interact.

\section*{Conflicts of Interest}
The authors declare no conflict of interest.

\section*{Acknowledgments}
No external funding was used to complete this work. ChatGPT 5.1/5.2 (OpenAI) and Grammarly were used for early brainstorming and as an editorial assistant related to organization, wording, clarity, and grammar in the text. These tools were not used to contribute to the core technical work, analyze the literature, or perform core technical analysis of the results. The authors have carefully reviewed and assume full responsibility for the final manuscript. All opinions and conclusions are solely those of the named authors and do not necessarily represent the positions or views of their institutions or the publisher of this work.  

\bibliographystyle{model1-num-names}
\begin{footnotesize}

\end{footnotesize}

\end{document}